# The Role of communication and network technologies in vehicular applications

**Yacine Khaled**,
*INRIA Rocquencourt - IMARA Team, France*
**Manabu Tsukada**,
*INRIA Rocquencourt - IMARA Team, France*
**José Santa**,
*University of Murcia - Computer Science Faculty, Spain*
**Thierry Ernst**,
*INRIA Rocquencourt - IMARA Team, France*

## ABSTRACT

Vehicular networks attract a lot of attention in the research world. Novel vehicular applications need a suitable communication channel in order to extend in-vehicle capabilities and, be aware about surrounding events. However, these networks present some proprieties, such as high mobility or specific topologies. These properties affect the performances of applications and more effort should be directed to identify the final necessities of the network. Few works deal with application requirements which should be considered when vehicular services are designed. In this chapter this gap is filled, proposing an analysis of application requirements which considers available technologies for physical/MAC and network layers. This study contains key factors which must be taken into account not only at the designing stage of the vehicular network, but also when applications are evaluated.

## INTRODUCTION

Nowadays, communications become essential in the information society. Everyone can get information anywhere, even in mobility environments, using different kinds of devices and communication technologies. In this frame the vehicle is another place where users stay for long periods. Thus, in addition to safety applications, considered as the most important services, other networked applications could bring an additional value for the comfort of drivers and passengers, as well as for driving efficiency, in terms of mobility, traffic fluency and environment preservation. However, such kinds of networks are characterized by a strong mobility, a high dynamicity of vehicles and specific topology patterns. Moreover, these networks experiment significant loss rates and very short communication periods. These properties affect the performance and feasibility of vehicular applications. The proper operation of vehicular applications remains a great challenge nowadays, and specific requirements should be considered. In our opinion, such kind of applications should be initially studied from the communication technologies points of view. An analysis of the requirements, in terms of technologies, should help to design efficient applications. In this chapter an analysis of available communication and network technologies is given, and a study about how they can fulfill main networking requirements of ITS applications is stated.

This chapter presents a study of application requirements in terms of communication technologies. First, broad background and our point of view are presented. In order to explain the main focus of our chapter, we started by describing vehicular applications and services, selecting, thus, the most representative ones as cases of study and defining their networking requirements. So, before analysing the application requirements against the capabilities offered by current communication possibilities, we introduce the most common communication technologies in the vehicular field as solutions issues, as well as and some of the most standardized level-three. Finally, some concluding remarks end the chapter.

## BACKGROUND

Numerous research works dealt with vehicular services, essentially for road safety purpose, but also for traffic efficiency and infotainment. However, the achievement of the functional goals of these applications is strongly linked to technological requirements, which vary from one application to another. For instance, safety applications should operate with good location accuracy, and real-time and scalable communications; distributed games or talk applications, however, do not need a great scalability or real-time features.

To ensure the appropriate operation of these applications regarding networking, new technological requirements, far away from traditional services of fixed networks, appear in vehicular communications. This kind of solutions usually needs to cover many networking necessities for their efficient operation in most cases. In this chapter, the most important ones are studied, and have been identified as: location awareness, geocast capabilities, penetration rate dependency, time awareness, permanent access and mobility.

To meet these demands, a number of communication technologies at access level are currently available. The most important ones are briefly introduced, such as Bluetooth, DSRC (IEEE 802.11p), cellular networks and satellite. Also, these technologies are analyzed according to communication paradigms covered, vehicle to vehicle (V2V), vehicle to infrastructure (V2I) or infrastructure to vehicle (I2V), and the destination nodes involved in the communication (1-to-1 or 1-to-n). In the same way, some of the main network (level-three) technologies studied in standardization bodies of vehicular communications are discussed, in order to determine which of the networking requirements can be covered with them. NEMO as well as some common MANET and VANET proposals are briefly described, but also more specialized concepts, like Multihoming, Flow distribution, Route Optimization and MANEMO, are analyzed. Finally, an overlay architecture using cellular networks shows the feasibility of this technology to enable vehicular communications.

## DESIGN OF EFFICIENT VEHICULAR APPLICATIONS

### Vehicular applications and services

Vehicular services and applications can be classified into three main families: safety, traffic management and monitoring, and comfort. Road safety certainly is the main motivation of most researchers and represents the major issue in Intelligent Transportation Systems (ITS). About 40.000 people died on roads every year only in the European Union, with around 1.7 million people incurring several injuries. These accidents are often caused by a faulty driver behavior, bad weather conditions or mechanical problems. Hence, one of the most important solutions relies on using vehicular communications to anticipate road accidents, extend road visibility and disseminate safety information. In addition to road safety applications, traffic information and monitoring systems are another important application field of vehicular networks. They aim at improving traffic flow and road usage, providing timely information about traffic state along many kilometers. Finally, the goal of comfort applications is offering novel on board services to improve the travel experience, improving common multimedia capabilities of current commercial vehicles.
In order to analyze this wide world of vehicular applications, some of the most representative ones have been chosen as cases of study. Thus, three reference applications which best represent these three families are described, such as Cooperative collision warning, Platooning or Parking place management.
Finally, a discussion about how networking requirements of vehicular application and services can be fulfilled by means of each technology is included in the chapter. Since there is no any specific technology which can satisfy all these requirements, our opinion is that future on board and road side units should combine some of them in order to enable the deployment of different ITS applications. Having the requirements and technologies in mind, vehicular network solutions can be designed by means of new communication architectures or integrating current ones. In this frame, our contribution should help in the design of both vehicular applications/services and technologies for testing scenarios, evaluations, commercial developments or the research field, in general.

### Safety applications

- Cooperative collision warning. It is considered as the most important safety application. It allows enhancing the driver capabilities by monitoring the distance between vehicles and, depending on the case, warning the driver or automatically breaking when the distance decreases under a threshold. These systems also take into account the post-collision situation, when vehicles on the road must be warned.
- Incident management. The aim of this system is to successfully manage current accidents on the road. First, by detecting road problems (e.g. obtaining location and nature of accident) via positioning devices or other sensors. The next point is to manage vehicle flows during and after the accident, through vehicular communications.
- Emergency video streaming. It deals with video forwarding in emergency contexts. Some vehicles are equipped with video cameras and have enough storage capabilities to buffer multimedia content. This service can be provided over V2V (vehicle to vehicle) communications [Guo, M-H., & ZeguraGuo et al.2005] or 3G [Qureshi, Carlisle, & GuttagQureshi et al.2006].

### Traffic management and monitoring systems

- Platooning. Such systems allow vehicles to travel together closely and safely. This leads to a reduction in the space used by vehicles on a highway. As a consequence, more vehicles can use the highway without provoking congestion. This kind of solutions increase comfort levels of passengers, and can allow a higher level of safety due to constant monitoring of the road state by the vehicles in the platoon. Although the operation of these systems requires, essentially, a direct communication between vehicles, some enhancements could be obtained by using other technologies, which can improve location accuracy or overcome the lack of V2V communication.
- Vehicle tracking. This service allows car manufacturers, logistic companies and other trusted parties, to remotely monitor vehicle statistics. Data is collected by an AU (Application Unit), and sent by the OBU (On Board Unit) to the data center through network technologies.
- Notification services. It consists of providing travel information to subscribers through an Internet access. After the subscription, a user can be notified when information is available. As application examples, we can quote weather and traffic forecasting.

### Comfort applications

- Parking place management. This service allows drivers to discover a free parking place and book it. Additionally, a vehicle could park itself without the need of driver assistance.
- Distributed games and/or talks. This kind of entertainment applications comprise the management of activities among a limited number of vehicles, in a distributed fashion and via a purely V2V link. For instance, we can quote card games, sharing draws or instantaneous talks.
- Peer to peer applications. Using these services it is possible to exchange data between vehicles, without contacting any application server. This exchange operates, essentially, by means of V2I communications and complementary through V2V. In this context, we have applications such as instantaneous messaging, file transfer and voice over IP.

The achievement of the functional goals of the previous applications is strongly linked to technological requirements, which vary from one application to another. For instance, safety applications should operate with good location accuracy, and real-time and scalable communications; distributed games or talk applications, however, do not need a great scalability or real-time features. Because of these reasons, the following section summarizes the most important application requirements, and use the applications described above as a reference point.

## Networking requirements of vehicular applications

In this section main technological requirements of vehicular applications and services, regarding networking, are carefully examined. This kind of solutions needs to cover many needs for their efficient operation in most of the cases. However, in our study we will treat only the most important ones.

### Location awareness

Next generation vehicles are expected to exchange information not only beyond their immediate surroundings and line-of-sight with other vehicles, but also with the road infrastructure and Internet databases. This will allow vehicles to anticipate trajectories, coordinate merging manoeuvres, notify a braking action to vehicles behind, warn oncoming traffic of an icy patch, report road traffic conditions, locate parking lots, or simply entertain passengers. In this context, the knowledge of their actual position and trajectory is necessary, and it is only meaningful to vehicles in a particular geographic area. The exchange of information among vehicles in a particular geographic area requires reliable and scalable communication capabilities, which we call geographical routing and addressing. This function mainly relies on the information given by GPS receivers. However, GPS imposes some constraints such as lack of coverage in some environments or its weak robustness for some critical applications. For these reasons, other positioning techniques such as cellular or WiFi localization, dead reckoning (by using last known position and velocity) [King, Füßler, Transier, & EffelsbergKing et al.2006], and image/video localization, have been proposed in the vehicular field [Boukerche, Oliveira, Nakamura, Jang, & LoureiroBoukerche et al.2008]. Critical safety services such as alert cooperative collision warning and incident management need a high accurate localization, as well as some comfort applications such as parking booking. Note that an accurate positioning system can help us to define the zone of relevance more precisely. Other services, however, require a low accurate localization, like peer to peer applications and vehicle tracking.

### Geocast capability

Geocast provides the capability to deliver a message to nodes within a geographical region [Maihöfer 2004]. The shape and size of this area depend on the application aims. The complexity of defining this region can be as high as the set of vehicles behind or in front of the subject one. Other times this constraint is relaxed, and defining this region as the vehicles inside a geographic area, or near a designated spot (such as a smog area), is enough. In order to advocate a general communication architecture, where services which require both unicast and geocast capabilities can be deployed, an hybrid networking architecture can be proposed [Khaled, Ducourthial, & Shawky.july, 2007]. This way, services such as platooning, which need unicast communications, do not experience bad performances. Geocast is considered efficient if the information is forwarded in both sparse and dense geographical areas, while efficiently leveraging the available bandwidth. This criterion, scalability, was introduced in [Clifford Neuman 1994], and it was defined as the ability to handle the addition of nodes or objects without suffering a noticeable loss in performance or increase in administrative complexity.

### Penetration rate dependency

Penetration rate is defined as the percentage of vehicles equipped with the necessary OBU on the road. This parameter may have important consequences in the operation of some applications [BREITENBERGER, GRBER, NEUHERZ, & KATES et al.2004], especially the critical and safety ones. Although a low penetration rate is obviously a problem in safety applications, such as collision avoidance, an excess of equipped vehicles also arises transmission problems. However, applications such as comfort do not have to be too much aware of this factor. In cellular networks, situations of high penetration are also a problem. The system performance is not affected when the number of equipped vehicles is low, but in high load circumstances, the network connection starts to give a poor performance when the time slot scheduler need to serve too much users [Landman & KritzingerLandman & Kritzinger2005]. Note that penetration rate has a direct bearing on the wireless bandwidth used. The higher the penetration rate, the higher the wireless bandwidth should be used to allow vehicles to communicate.

### Time awareness

Vehicular applications often require a reliable communication channel that supports time-critical message transmissions [Meier, Hughes, Cunningham, & Cahill et al.2005]. One of the most important criterions for measuring the quality of the network, regardless of the application type, is the communication delay. Although most applications have time constraints, those related with road safety are critical. Due to this, a challenge in vehicular networks is providing a real-time behavior. In order to enable the driver to react quickly, the information must reach the destination in a very small delay following the event. However, this requirement is not easy to ensure in mobile networks. This difficulty is even greater if we consider vehicular network characteristics, particularly, the high mobility. Thus, real-time communications can only be assured by the presence of an efficient and robust communication system.

### Permanent access

Permanent access to the network is one of the main drawbacks of vehicular communications. In VANET designs, a physical infrastructure is not necessary, due to the inherent decentralized design. Regarding infrastructure-based networks, operators do not offer the same service over the entire terrestrial surface. For instance, over urban environments, the coverage is excellent, and the amount of base stations where the mobile terminal could be connected is really high. At rural locations, however, the deployment is poor. A vehicle equipped with a VANET system, however, is always able to emit messages because the vehicle itself is part of the infrastructure. Moreover, in cellular network connections, it is also important to differentiate between two important concepts regarding the access to the network: coverage and capacity. The coverage can be understood as the possibility of the mobile terminal to use the network, because at a particular location operators have deployed the necessary infrastructure. However, the user can be rejected to establish a call or a data connection, even in good coverage circumstances, if the capacity of the network has been exceeded. Depending on several technological issues, such as modulation, frequency allocation, time slot scheduling, etc., this effect has a different impact. This way, the number of users who are concurrently using the network restricts the potential cellular network usage. At the application level, some services such as file transfer or download need a permanent communication channel. In this kind of applications, the election of a suited vehicular network is essential.

### Mobility

Wireless network technologies allow devices to move freely. However, this mobility affects the potential permanent access to the network (see the previous point) and causes other problems. In [Wewetzer, Caliskan, Meier, & Luebke et al.2007], experimental evaluations give real results of these effects. In 802.11 transmissions the distance between the sender and receiver is an important factor; the more the distance, the smaller the probability of reception of packets, as also show [ElBatt, Goel, Holland, Krishnan, & Parikh et al.2006, Khaled, Ducourthial, & Shawky et al.2005-Spring]. In infrastructure-based technologies, handoffs between base stations are also relevant, due to the potential decrease of performance in the process. Poor latency and throughput results are obtained if the mobile terminal is moving at locations far away from the UMTS Node B without performing a handoff [Alexiou, Bouras, & Igglesis et al.2004]. Nevertheless, the distance between two devices during the communication is not the only noticeable effect of mobility. Interference with other radio equipments in the case of VANET should also be taken into account, due to the wide usage of the 2.4 GHz frequency band [Wewetzer, Caliskan, Meier, & Luebke et al.2007]. The presence of the equipment at locations of bad orography could also cause communication problems in vehicular networks. Other external factors, like the existence of other vehicles or buildings are considered in realistic mobility patterns for VANET solutions [Naumov, Baumann, & Gross et al.2006]. The knowledge of both requirements and application allow us to efficiently identify the needs of each application. Thus, we use the applications introduced in the previous section to evaluate their requirements. This evaluation study is summarised in Table 1.

| Applications/App. Req. | Location awareness | Geocast capability | Penetration rate dependence | Time Awareness | Permanent access | Mobility |
|---|---|---|---|---|---|---|
| **Safety** | | | | | | |
| Cooperative Collision Warning | ★★ | ★★ | ★★ | ★★ | ★ | ★★ |
| Incident management | ★★ | ★★ | ★★ | ★★ | ★ | ★★ |
| Emergency video streaming | ★★ | ★★ | ★★ | ★★ | ★ | ★★ |
| **Traffic management and Monitoring** | | | | | | |
| Platooning | ★★ | ★★ | ★★ | ★★ | ★ | ★★ |
| Vehicles tracking | ★ | | ★ | ★ | ★★ | ★ |
| **Notification Services** | ★ | | ★ | ★ | ★ | ★ |
| **Comfort** | | | | | | |
| **Parking place management** | ★★ | ★★ | ★ | ★ | ★ | ★★ |
| **Distributed games and/or talk** | ★ | ★★ | ★★ | ★ | ★ | ★★ |
| **Peer-to-peer** | ★ | | ★ | ★ | ★★ | ★ |

*Table 1: Application requirements (none-not needed, ★ needed, and ★ suited this requirement)*

## Communication technologies

Wireless communication technologies are increasing nowadays, with the aim of substituting typical wired connections and improve mobility. At the same time, the vehicular field is currently introducing into the telematics world, where informatics and telecommunications try to improve traffic security, efficiency and safety. In this frame, wireless communications are essential to connect the vehicle with the environment.

### *Bluetooth*

Bluetooth is a wireless standard (802.15.1) specially created for short range communications between devices usually connected by local ports. Thanks to Bluetooth, however, it is possible to create a personal area network (PAN) where several devices can be connected. It operates in the 2.4 GHz band and, due to the low power consumption features, allow communications in a typical range of tens of meters. Bluetooth terminals are grouped in piconets, and these piconets can also be connected by means of scatternets.

The properties of Bluetooth make it perfect for in-vehicle networks [Nolte, Hansson, & Lo Bello et al.2005]. Some researchers also advocate the usage of Bluetooth for V2V applications [Sugiura & Dermawan & Dermawan2005]. However, this technology is limited by the necessary time to form piconets and scatternets (in the order of seconds) [Sawant, Tan, Yang, & Wang et al.2004] and, overall, the limited communication range.

### *WLAN and DSRC*

Wireless Local Area Networks (WLAN) were created to cover connectivity requirements usually fulfilled by common LAN technologies, like Ethernet. The set of standards which deal with WLAN features are inside the 802.11x group, and consider a set of protocols which allow terminals to be connected to a base station, which is in charge of connecting computers to the rest of the wired network. Among these standards, 802.11a/b/g specifications are the most known. IEEE 802.11a was the first adopted WLAN technology, offering a maximum rate of 54 Mbps over distances of 100 meters. However, the used 5 GHz band is not available (mainly) in some European countries, and the 802.11b standard was finally accepted as the definitive WLAN technology. 802.11b implements the same core protocols than 802.11a, but uses the 2.4 GHz band, what decreases absorption problems due to walls and other obstacles. This way, the communication range is augmented until 140 meters, but data bandwidth is maintained under 11 Mbps. 802.11g overcome bandwidth limitations of 802.11b with a new modulation scheme over the 2.4 GHz, presenting the successor of 802.11b. Although the most common usage of 802.11 technologies is the infrastructure mode, using a base station, these devices can also be configured to directly communicate with another terminal, using the ad-hoc mode.

This one is preferred to enable vehicular communications. Many V2V works use WLAN technologies to test multitude of applications, such as cooperative collision avoidance using V2V communications between nearby vehicles [Ueki, Tasaka, Hatta, & Okada et al.2005, Ammoun, Nashashibi, & Laurgeau et al.2006], or multi-hop strategies [Biswas, Tatchikou, & Dion et al.2006]. However, common WLAN standards have some limitations when critical information has to be transmitted in the vehicular environment [Yousefi, Bastani,&Fathy et al.2007]. For this reason, USA, Japan and Europe have allocated a specific band in the 5.8 and 5.9 GHz for vehicular transmissions, using Dedicated Short Range Communications (DSRC). A variation of the 802.11 standards, 802.11p, is being used as background in the DSRC research. This standard covers the requirements for communicating both periodic and critical information, which allows the deployment of a great variety of vehicular services, using both V2V [ElBatt, Goel, Holland, Krishnan, & Parikh et al.2006] and vehicle to road side communications [Hattori, Ono, Nishiyama, & Horiuchi et al.2004]. Take notice that 802.11x could be considered as the most popular technologies.

### *Cellular networks*

Since initial analog technologies, such as the American AMPS, cellular networks have been gradually improved in terms not only of availability all around the world, but also in the quality of service offered. As a result of applying digital communications to cellular networks, the GSM (Global System for Mobile communications) technology achieves the purpose of spreading mobile phones among normal population. Its wide adoption in Europe last years has led the expansion of GSM to other potential markets, like the Chinese one. Many people usually identify the GSM technology as the second generation (2G) of cellular networks, which substituted the first one, based on analog technologies. Although the main concern of cellular networks, until some years ago, was focused on telephony purposes, data connections are becoming more and more popular these days. GPRS (General Packet Radio Service) appeared with the aim of providing higher data rates than the 9.6 Kbps offered by the standard GSM. GPRS provides a maximum of 177/118 Kbps in the downlink/uplink channels, and it is understood as the intermediate step between 2G and 3G, hence this is the reason why it is called 2.5G. Last years, the expansion of CDMA (Code Division Multiple Access) communication technologies has lead to the appearance of the 3G cellular networks. CDMA2000 and UMTS (Universal Mobile Telecommunications System), this one as the evolution of GSM 2G, are two of the most extended 3G technologies. UMTS offer 384/128 Kpbs, but the recent HSPA (High Speed Packet Access) improvements offer maximum data rates of 14.4/11.5 Mbps.

The introduction of cellular networks in the vehicular domain comes from several years ago, when GSM or GPRS data connections started to be used in tracking and monitoring systems. The appearance of GPRS also made possible the usage of cellular networks for providing traffic information or emergency warnings [Masini, Fontana, & Verdone et al.2004]. However, until the arrival of 3G technologies, low data rates had avoided the spread of cellular networks in ITS [Adrisano, Verdone, et al.2000]. The advantages of the UMTS communication medium in mobility environment are defended by some authors, which use the UMTS aerial interface for direct V2V communications. The usage of the UMTS operator's infrastructure in bidirectional communications is present in the literature, as monitoring systems [Hoh, Gruteser, Xiong, & Alrabady et al.2006] for example, but its application for V2V communications is still a challenge, due to inherent delay problems. Another drawback of using data connections with cellular networks is the extra money which has to be paid for the usage of the operator's infrastructure. Current trend is paying a fixed quote per month, with an extra cost if the transmission rates fall out of the contract, but it is expected that the adoption of UMTS among the population and the vehicular field decrease the price of the final bill, by means of special agreements with operators ["3G/UMTS Evolution: towards a new generation of broadband mobile services"]. Apart of this, some people think that a general communication technology for the ITS domain is still needed, and cellular networks could be the solution [Kiess, Rybicki, & Mauve et al.2007].

### *WiMAX*

WiMAX, or Worldwide Interoperability for Microwave Access, is a communication technology which try to fill the gap between 3G and WLAN standards, and it is the first implementation which appears to comply with theMAN(Metropolitan Area Network) concept, in a wireless manner. Two main standards are currently considered: 802.16d and 802.16e. The first one is used at fixed locations, and it is a perfect solution for connecting different buildings of a company at a low cost, for example. This specification offers up to 48 Km of coverage and data rates of 70 Mbps. The 802.16e standard, specifically designed for mobile users connected to a base station. The OFDM (Orthogonal Frequency Division Multiple access) technology is used in this standard to serve multiple users, and the final physical interface considered copes with mobility issues, such as interferences, multipath and delays. 802.16e is, hence, the most appropriate specification of WiMAX for the vehicular field. Tens of Mbps, mobility speed up to 100 Km/h, and 10 km of coverage to the base station, make 802.16e a good option for urban scenarios, where vehicles can be connected at a high data rate using a WiMAX deployment. Currently it is possible to obtain some Pre-WiMAX devices, but it is expected that, as soon as the final specifications are ready, the spectrum of vehicular services which could be deployed with WiMAX grow rapidly. In [Han, Moon, Lee, Jang, & Lee et al.2008] the authors analyze the performance of WiMAX in a subway, where the maximum speed is 90 Km/h. The results show that a mean of 2 Mbps and 5.3 Mbps can be obtained in real scenarios, with an average RTT of 100 ms. Venturi has designed an electrical vehicle which uses a pre-WiMAX interface for remote monitoring purposes, as a joint work between the vehicle manufacturer and Intel Corporation.

### *RDS and TMC*

The Radio Data System (or RDS) was developed to carry digital data using the common FM radio band. This allows to multiplex additional information with the audio emission, such as the name of the radio station or the current song, but also it can include a data flag which indicates the receiver it has to pay attention to the broadcasting information because it is being transmitted a traffic bulletin. RDS offers a data rate of 1187.5 bps, and the transmission range offered by FM can reach locations at 80 kilometers far way. The RDS version deployed in U.S. is called RBDS (Radio Broadcast Data System) and operates almost identically as RDS, however its usage is less common.

A more suitable solution for traffic information dissemination is offered, however, by the Traffic Message Channel (TMC) system. With this system, information about traffic problems is broadcasted digitally, so an appropriate navigation device can warn the user and calculate an alternative route, for instance. The notifications reported by TMC include an event identifier and the location of the problem. TMC traffic is usually transmitted through RDS, and this is the reason why both technologies are usually put together.

### *Satellite*

Satellite communication consists of three main entities: sender station, satellite system, and receiver devices. First of all, data is sent from the sender station to the satellite, which is in charge of forwarding the information to receiver devices.

Satellite communications offer a very wide coverage and great broadcast capabilities. It is suited to provide connectivity at remote places, such as mountain areas or islands, but also in developing countries. The data can be sent from an only sender to multiple receivers at the same time and using the same frequency. Thus, satellite communications are suitable for multimedia broadcasting, such as live video, movies and music.

Although sender stations and receiver devices are usually installed at fixed locations, the later ones can be mobile and equipped in vehicles. This kind of architecture is feasible for a unidirectional system providing an I2V service; however it must be taken into account the important delay which suffer data packets, due to the propagation distance to and from the satellites. The bandwidth obtained in a mobile terminal is between 300 and 500 kbps. A sender station is usually too big to be brought inside a vehicle, and it requires a precise orientation to the satellite used. The UniDirectional Link Routing (UDLR) [Duros, Dabbous, Izumiyama, Fujii, & Zhang et al.2001] has been standardized to emulate bidirectional communications with a satellite unidirectional link, where mobile terminals receive data using the satellite channel and transmit using other access technologies.

### *Synthesis*

Connectivity necessities of vehicles can be divided in two main groups: vehicle to vehicle communications (V2V, VVC) or inter-vehicle communications (IVC), and communications with the infrastructure. In the literature, many authors use vehicle to infrastructure communications (V2I) to denote both data flow directions, however, according to the specific use of several technologies for one or the other communication pattern, it is more correct to distinguish between V2I and I2V (infrastructure to vehicle communications). It is important to consider this whole set of communication possibilities for vehicles because, depending on the application or service necessities, we will have to decide among one of the available wireless network technologies.

Apart from the communication pattern covered, wireless communication technologies can be divided into those which establish 1-to-1 physical links, and those which consider 1-to-n broadcast ones. In this last case, some kind of access point is in charge of sharing out the available bandwidth among the clients. This bandwidth, thus, could become insufficient when the number of served nodes increases inside the coverage area. Due to this, the tendency in short-range wireless technologies is taking advantage of the available bandwidth, sharing it among a small number of users because, anyway, the coverage is small. On the contrary, wide-range technologies must share the available bandwidth among much more users. However, short-range wireless media lack on stability, due to the small accessible area. It is also important to remark how V2V communications are obtained by means of 1-to-1 technologies, and communications with the infrastructure are commonly created using the 1-to-n ones.

A brief overview of main wireless technologies used in the vehicular domain is given in Table 2. For communications with the infrastructure, it is said that WLAN, DSRC, WiMAX, cellular and satellite are feasible. However it is important to remark the different application they cover in this domain. In the case of WLAN/DSRC, vehicles usually connect with local roadside units, what usually is called vehicle to road side communication. On the other hand, in the WiMAX/cellular case it is used a medium range 1-to-n network, and in satellite communications a wide range 1-to-n model is applied. In the cellular and satellite cases the design of the network is even more fixed than with any other technology, because we have to use the operator's installations. This way, service providers usually consider the direct Internet connection offered by the operator, and there is no possibility to manage data traffic inside the operator's network.

| Technology | Range | Link type | Data rate | Frequency band | Standard | Standard Vehicular applicability | | |
|---|---|---|---|---|---|---|---|---|
| | | | | | | V2V | V2I | I2V |
| **Bluetooth** | 100 m | 1-to-n | 1 Mbps | 2.4 Ghz | IEEE 802.15.1 | ★ | | |
| **WLAN** | 200 m | 1-to-1<br>1-to-n | 10-50 Mbps | 2.45 Ghz | IEEE 802.11a/b/g | ★★ | ★ | ★ |
| **DSRC** | 1 Km | 1-to-1 | 50 Mbps | 5.9 Ghz | IEEE 802.11p | ★★ | ★★ | ★★ |
| **WiMAX** | 10 Km | 1-to-n | ~20 Mbps | 2.45 Ghz | IEEE 802.16e | | ★★ | ★★ |
| **Cellular** | 10 Km | 1-to-n | ~10 Mbps | 700-2600 Mhz | n/a | | ★★ | ★★ |
| **RDS/TMC** | 80 Km | 1-to-n | 1187.5 bps | 87.5-108.0 Mhz | CENELEC EN 50067<br>CEN ENV 12313 | | | ★★ |
| **Satellite** | >10.000 Km | 1-to-n | 300-500 Kbps | 950-1450 Mhz | n/a | | ★ | ★★ |

*Table 2: Properties of main vehicular communication technologies, and vehicular applicability (none-not possible, ★ possible, and ★ suited)*

## Setting up the communication channel

In this section, the main network (level-three) technologies treated in standardization bodies for vehicular communications are described. NEMO, MANET and VANET are briefly described. Moreover, more specialized concepts, like Multihoming, Flow distribution, Route Optimization and MANEMO, are included. Fig. 1 shows a vehicular networking scheme where the most important technologies at level-three are included in an integral communication solution. Finally, an overlay architecture using cellular networks shows the feasibility of this technology to enable vehicular communications.

### *NEMO*

The NEMO Basic Support [Devarapalli, Wakikawa, Petrescu, & Thubert et al.2005] functionalities involve a router on the Internet to allow mobile computers to maintain a global connectivity to Internet. In the ITS field, the basic scheme is represented in Fig. 1, and is described as follows. A Mobile Router (MR) located in the vehicle acts as a gateway for the Mobile Network of the vehicle, and manages mobility on behalf of its Mobile Network Nodes (MNN). The MR and a fixed router in the Internet, called Home Agent (HA), establish a bi-directional tunnel which is used to transmit the packets between the MNN and their Correspondent Nodes (CN). In vehicular networks, this mechanism is often referred as a vehicle to infrastructure (V2I) communication pattern, because it involves the transmission of information through the fixed Internet.

Notice that in this scheme, the OBU can act as MR and the AU can be considered as a generic MNN. In the latter case, RSUs are attachment points either acting themselves as IPv6 access routers or as bridges directly connected to an access router.

### *Multihoming*

MRs can be shipped with multiple network interfaces such as IEEE802.11a/b/g, WiMAX, GPRS/UMTS, etc. When a MR maintains these interfaces simultaneously up and has multiple paths to the Internet, it is said to be multihomed. In mobile environments, MRs often suffer from scarce bandwidth, frequent link failures and limited coverage. Multihoming comprises some benefits to alleviate these issues [Ernst, Montavont, Wakikawa, Ng, & Kuladinithi et al.2007]. The possible configurations offered by NEMO are classified in [Ng, Ernst, Paik, & Bagnulo, et al.2007], according to three parameters: (x) the number of MRs in the mobile network, (y) the number of HAs serving the mobile network, and (z) the number of MNPs (Mobile Network Prefixes) advertised in the mobile network. NEMO basic support has a “single MR, single HA and single MNP” configuration, referred to as (x; y; z) = (1; 1; 1). In this configuration, a tunnel is established between the HA address and a Care-of Address (CoA) of the MR in NEMO Basic Support, even if the MR is equipped with several interfaces. Multiple Care-of Addresses Registration (MCoA) [Wakikawa, Ernst, Nagami, & Devarapalli et al.2008] is thus proposed as an extension of both Mobile IPv6 and NEMO Basic Support to establish multiple tunnels between MR and HA. Each tunnel is distinguished by its Binding Identification number (BID). In other words, NEMO Basic Support only realises interface switching, while MCoA supports simultaneous use of multiple interfaces.

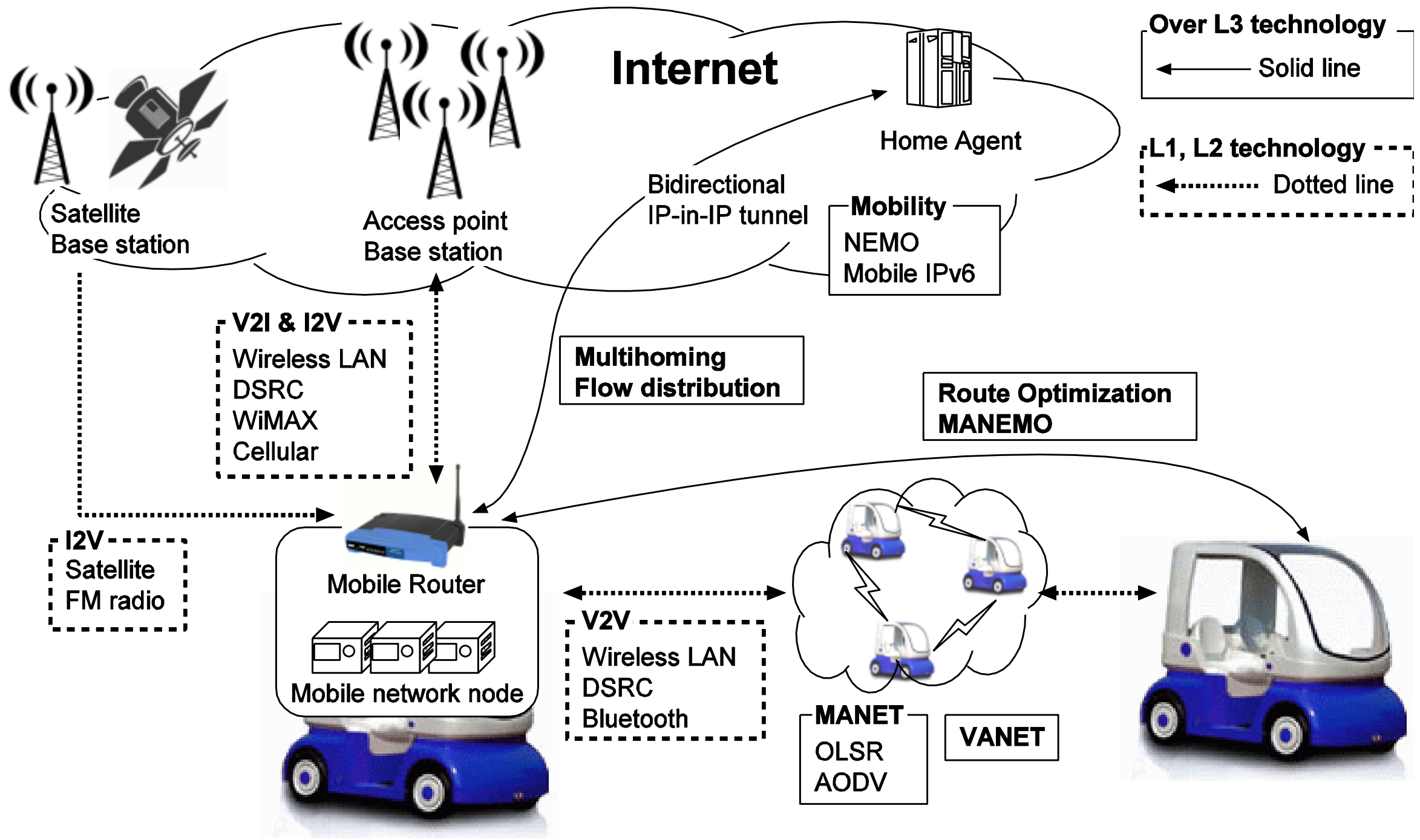

*Figure 1: Overview of the application of network technologies in the vehicle domain*

### *Flow distribution*

To transfer data through multiple interfaces, a policy based flow distribution mechanism is used. The traffic can be distributed by multiple paths considering the source and destination addresses, source and destination ports, flow type, and so on. In NEMO basic support, traffic from the Internet to the mobile network is distributed by the HA, while the distribution in the opposite direction is carried out at the MR. This way, neither MR nor HA are able to change the complete round-trip path. Charging this operation through policy rules can provoke, however, asymmetric paths which could not satisfy the user's demands. For this reason, a policy synchronisation method between MR and HA is needed. Some proposals have been considered at the IETF [Soliman, Montavont, Fikouras, & Kuladinithi et al.2007, Larsson, Eriksson, Mitsuya, Tasaka, & Kuntz et al.2008].

### *Route Optimization*

NEMO is one of the main level-three technologies of vehicle communication, however, some issues related to Route Optimization still remain unsolved in NEMO Basic Support, while they have already been covered in Mobile IPv6 [Johnson, Perkins, & Arkko et al.2004].

In NEMO, all the packets to and from MNNs must be encapsulated and sent by means of an IPs tunnel between the MR and the HA. Thus, all these packets between MNNs and CNs must go through the HA. This arises several performance issues.

Suboptimal routes are caused by the mandatory pass of packets through the HA. This leads to increased delays, undesirable for applications such as real-time multimedia streaming. Packet Encapsulation implies an additional head of 40-bytes, which can cause packet fragmentation. This also results in an increased processing delay at the encapsulating and decapsulating stages in both MR and HA, respectively. Bottlenecks in the HA are an important issue, because traffic to and from MNNs is aggregated at the HA when it supports several MRs acting as gateways for several MNNs. This may cause congestion at the HA, which could lead to additional packet delays, or even packet losses. Nested Mobile Networks is an issue that NEMO Basic Support raises. This permits a MR to host other MRs inside the mobile network. With nested mobile networks, the use of NEMO further amplifies the sub-optimality previously described. In IETF, route optimization issues of NEMO are addressed in [Ng, Thubert, Watari, & Zhao, et al.2007]. Requirements of route optimization in various scenarios are described for vehicular networks in [Baldessari, Festag, & Lenardi et al.2007], and for aeronautic environments in [Eddy, Ivancic, & Davis et al.2007].

### *MANEMO*

Both MANET and NEMO have been designed independently as layer-three technologies. NEMO has been designed to provide global connectivity, and MANET to offer direct routing in localised networks. MANEMO comprises the usage of both concepts, MANET and NEMO, together, which could bring benefits for route optimization. Since direct routes are available in MANET, it can provide direct paths between vehicles, as Fig. 1 shows. These paths are optimized and tunnel-free, reducing overhead [Wakikawa, Okada, Koodli, Nilsson, & Murai et al.2005, J. Lorchat and K. Uehara 2006, Tsukada, Mehani, & Ernst et al.2008]. One possible topology configuration using MANEMO is described in [Wakikawa, Clausen, McCarthy, & Petrescu et al.2007], and issues and requirements of such architectures are summarised in [Wakikawa, Thubert, Boot, Bound, & McCarthy et al.2007]. In addition, MANEMO is also used in vehicular communications, for example, VARON [Bernardos, Soto, Calder´on, Boavida, & Azcorra et al.2007] focuses on NEMO route optimization using MANET. It also provides the same level of security as the current Internet, even if the communication is done via the MANET route.

### *MANET and VANET*

Mobile Ad hoc Networks (MANET) are suitable for wireless routing applications within dynamic topologies. This type of communication does not require any infrastructure. In order to route messages in such a network, each node is invited to participate in the message forwarding. Vehicular Ad hoc NETworks (VANET), a particular case of MANET, are characterized by a strong mobility of the nodes, a high dynamic topology, a significant loss rate, and a very short duration of communication. In these networks the node location is never stable, either locally or globally, and routing messages is a great challenge.

Many works have been done to design ad hoc routing algorithms to deal with the node's mobility: periodically updating routing tables by means of proactive algorithms (e.g. OLSR [Clausen et al.2001]); discovering routes under demand by means of reactive algorithms (.e.g AODV [Chakeres & Belding Royer 2004]); using geographical information to improve routing (e.g. GAMER, LBM [Maihöfer 2004], GPSR [Mauve, Widmer, & Hartenstein et al.2001]); detecting stable structures, or clusters [Jiang, Li, & Tay et al.2001]; using the node's movement for transporting messages [Zhao & Cao 2006]; following a broadcast approach for messages forwarding [Alshaer & Horlait 2005], etc. Some other protocols try to send packets only to aset of nodes located in a geographical zone (geocast), such as GeoGRID [Maihöfer 2004], for example. Here, the geographic area is divided in 2D logical grids. In each grid, one node is elected as the gateway, and only this one is allowed to forward messages.

### *P2P overlay network over cellular networks*

The usefulness of cellular networks in an architecture which allows communications between vehicles and with the infrastructure is presented in [Santa & Gomez- Skarmeta 2008]. The network architecture uses a P2P approach over the cellular network basis to enable vehicles to receive and send data packets.

Fig. 2 shows a general diagram of the proposed communication architecture. Traffic zones are organised in coverage areas, each one using different P2P communication groups. These zones are logical areas which do not have to fit in the cellular network cells. Information about the geometry of each area is maintained in the Group Server entity, and vehicles are able to move from one P2P group to another through a roaming process between coverage areas. This roaming is based on the vehicle location, provided by the GPS sensor. Information about areas is received from the Group Server using a TCP/IP link over UMTS. A local element called Environment Server manages special messages inside the area. These data packets are sent and received by service edges, located either at the vehicle or at the road side (Environment Servers). Messages are encapsulated in JXTA frames which are finally sent as UDP packets.

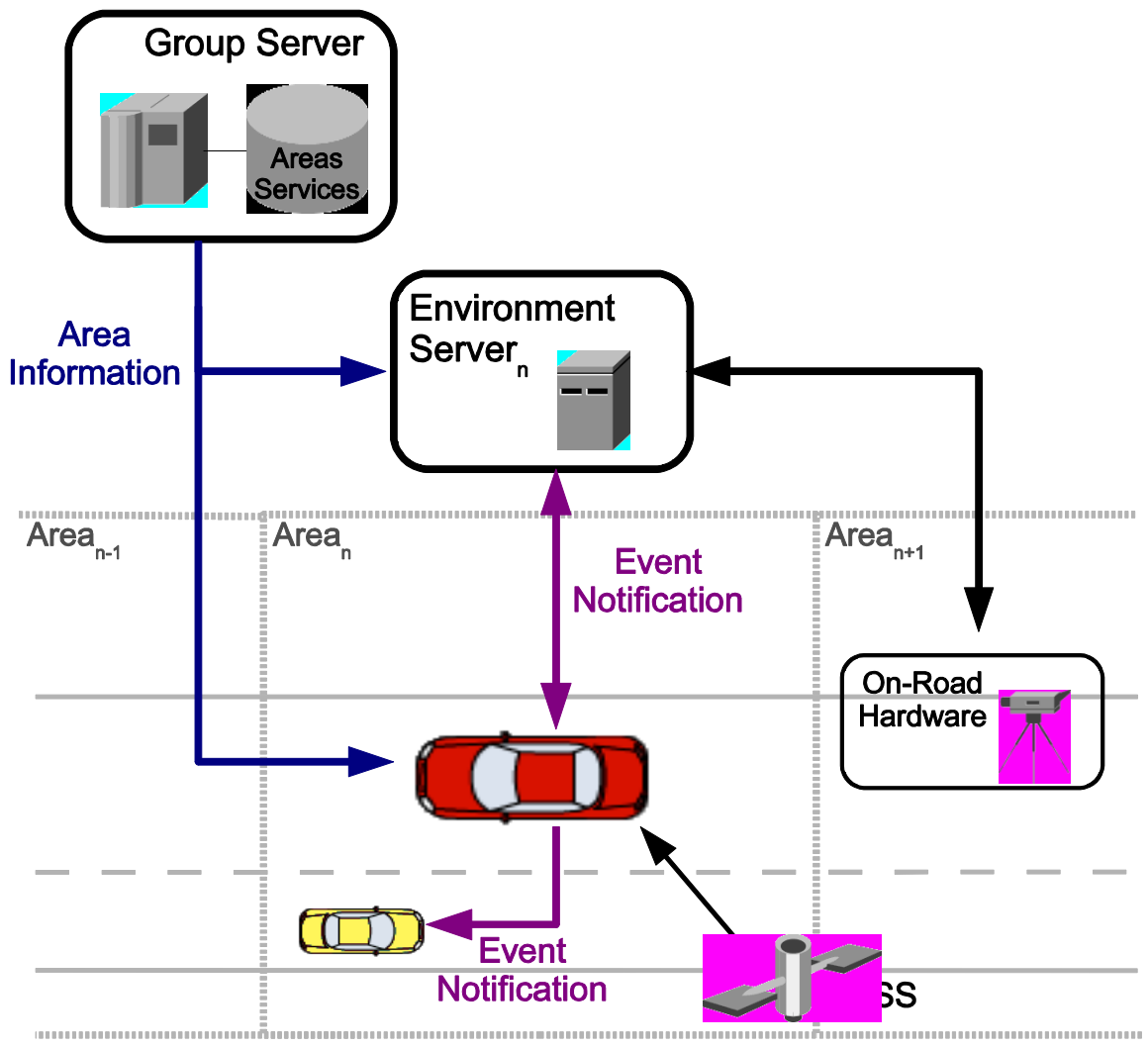

*Figure 2: P2P/Cellular network overlay network for vehicular communications*

## APPLICATIONS REQUIREMENTS ANALYSIS AND FUTURE RESEARCH DIRECTIONS

In this section, we analyze the previously presented vehicular applications requirements regarding the communication and network technologies points of view. First, we start by explaining the contribution of layer-one/two technologies on applications requirements. Then, with the same aim, we discuss the layer-three technologies role.

Regarding wider range communication technologies, such as satellite, RDS, cellular and WiMAX, some common aspects can be found considering the application requirements they can fulfill. All of them can avoid penetration rate problems present in short range technologies (e.g. WLAN, Bluetooth and DSRC). This is solved by the necessary infrastructure which provides the access to the network. This fact also helps to enable the provision of an almost permanent access to the network with high degrees of mobility. However, the amount of users simultaneously connected to the network is limited by the available bandwidth, which has to be shared among all of them. RDS, however, does not present this problem, because it only permits traffic in the downlink channel, as it is also noticeable in most of satellite deployments. Regarding geocasting, all wide-range technologies can simulate this feature by means of base station broadcasting. However, the performance of this method is limited by the size of coverage areas, which sometimes can be too large, as occurs in the extreme case of satellite communications. An overlay network like the one presented in [56] can solve this problem. Location functionalities are possible in cellular and WiMAX networks, following a detection mechanism at the base station. Cellular networks take advantage of this method in most urban environments to provide a good approximation of the user position. Finally, it is important to treat the delay in these wide-range technologies. Real time services can also be considered using cellular and, overall, WiMAX. However, in the case of safety services, the infrastructure available in the region of interest has to be evaluated.

WLAN takes advantage of its two communication modes, infrastructure and ad hoc, to ensure a permanent access under the lack of one of them. Also, in the same way, it can help to improve the penetration rate dependency by combining these two communication modes. In infrastructure mode, access points can contribute to improve the location accuracy and provide geocast capabilities, as it is noticeable in the GeoNet Project [GeoNet Project]. DSRC, or specifically IEEE 802.11p, allows the proper functioning of critical applications such as safety services, considering real-time constraints. This is carried out by using a specific emergency channel and some priority levels for the traffic. Although Bluetooth technology is limited by its range and connection time, it can be used in some situations, for instance when vehicles are very close, to maintain a permanent access to the network and alleviate the penetration rate problem.
Starting with level-three considerations, a special mention should be made to VANET protocols. According to vehicular network characteristics, high mobility and dynamicity, the combination of MANET concepts with location is found as a good solution in most cases. Thus, many position-based routing solutions have been proposed (e.g. GPSR [Mauve, Widmer, & Hartenstein et al.2001], DREAM [Mauve, Widmer, & Hartenstein et al.2001]), integrating location information in the routing messages. Note that some of these works have been introduced in standardisation bodies (like ETSI [ETSIs. d.]]), vehicular consortiums (C2C-CC [Car-to-Car Communication Consortium]), and European projects (GeoNet). The geographical information about the vehicle could also be used to perform geocast routing, where the messages are forwarded to a defined geographic area. As explained previously, V2V communications are highly dependent on the penetration rate. It is difficult to ensure a good communication under low and high load. To reduce this dependence, some solutions have been proposed. For example, in high penetration conditions, Geocast approaches can use the directed flooding approach, where the closer vehicle to the destination retransmits the message (e.g. LBM [Maihöfer 2004]). Under low load conditions, some solutions have been proposed, where the messages progress towards the destination by means of node movements [Kosch 2002, Zhao & Cao 2006, Li & Rus 2000, Chatzigiannakis, Nikoletseas, & Spirakis et al.2001]. These solutions offer real-time features, by optimizing communications. For applications that require a permanent access, VANET can be seen as an important asset in places where infrastructure-based networks are not available. Although location information is still lacking in IPv6 (or NEMO), many solutions and project deal with the geographical location. First, IPv6 can be implemented over geographical routing [Baldessari,Festag, Matos, Santos, & Aguiar et al.2006, Baldessari, Festag, Zhang, & Le et al.2008, GeoNet Project. d.], thereby use this information for routing and geocast communication. Other approach consists of extending IPv6 with geographical information [Chen, Steenstra, & K-S et al.2008, Vare, Syrjarinne, & Virtanen et al.2004].
The maintenance of multiple network interfaces (e.g. WLAN, WiMAX, GPRS/UMTS), thanks to multihoming, allows a permanent access to the network, regardless of the communication technologies. For instance, WiMAX or cellular could be used when WLAN is not available, and vice versa. The traffic distribution into multiple paths can increase the available bandwidth considerably and decrease the communication delay. Route optimization also improves the communication quality by reducing transmission delay. Moreover, the combination of NEMO and MANET takes advantage of both technologies, offering a continuous access to the network assuring the permanent access capability. The achievement of the applications requirements by using communication and network technologies is summarized in Table 3. For instance, we can notice that VANET communication could ensure the location and time awareness, and geocast capability. Also, NEMO could enhance the mobility and ensure a permanent Internet access. Due to this, we can notice that VANET and NEMO are very complementary in order to fill these requirements.

This analysis could be used as starting point for future research works. Thus, the main purpose of this study is to select and perform efficiently the appropriate communication technologies. However, the performances of these technologies are needed in order to know their feasibility, where the importance of this study in real environment. The latter consideration also helps an efficient integration, either simultaneously or separately, of communication technologies of different communication technologies and network proposals.
Take notice that, in our analysis study, some open research issues could be considered regarding the applications requirements. Among these issues, geocasting feasibility with both short and wide-range technologies could be exhaustively studied. However, about the integration of geographical information into the routing and addressing mechanisms, it is not clear which approach is adapted for which application, and what's impact of these approaches. So, additional studies are required.
The security is one of the main consequences of the vehicular networks proprieties. Thus, all the security issues should be reconsidered and adapted for VANET. Although the research activities are very important, the standardization issue for accelerating deployment of these applications by considering their requirements still crucial. This, without forgotten the support of public administration to design novel vehicular services.
Although some communication architectures have been proposed, this latter should keep its modules independent from technologies viewpoints. So, the modules need to be Interchangeable; for example 802.11b can be replaced by 802.11p in the future. In order to extend the usage of vehicular applications and meet user's demands, the industry should involve developing a new device like for cellular networks. Also agreements between services providers and network operators to enable the deployment of telematics services with a lower communication cost are highly needed.

| Technology/App. Req. | Location Awareness | Geocast capability | Penetration rate dependence | Time Awareness | Permanent Access | Mobility |
|---|---|---|---|---|---|---|
| **Layer 1/ 2** | | | | | | |
| **Bluetooth** | Possible | | Sensitive | | Help | O |
| **WLAN** | Possible | Help | Sensitive | Possible | Help | O |
| **DSRC** | Possible | Help | Sensitive | Possible | Help | O |
| **WiMAX** | Possible | Help | | | Help | O |
| **Cellular** | Possible | Help | | | Help | O |
| **DS/TMC** | X | | | | Help | O |
| **Satellite** | X | | | | Help | O |
| **Layer3** | | | | | | |
| **VANET** | O | O | Sensitive | O | Help | O |
| **NEMO** | | | | | O | O |
| **Multihoming** | | | | | O | |
| **Route optimization** | | | | O | | |
| **MANEMO** | | | | O | O | O |

**O** = The technology fills the requirement (The requirement is the aim of the technology).
**X** = The technology cannot fill the requirement.
**Possible** = The technology can be used to fill the requirement with other technologies (The requirement cannot be fill by only the technologies)
**Sensitive** = Penetration rate dependency sensitive (Note: no technology can increase penetration rate, people equips it because it is cheap or useful, for example.)
**Help** = The technology help the requirement (The requirement cannot be filled by only the technology).

*Table 3: Application requirements fulfilled with each technology*

## CONCLUSION

Many vehicular applications and services have been proposed at the beginning of this chapter. However, in order to achieve good performances, they should take into account the communication technology used. In this chapter we give an analysis of the available communication technologies, and study how they can fulfill the main networking requirements of ITS applications.

The initial overview of application and services sorts them into three main families: safety, traffic management and monitoring, and comfort. Then, we describe three reference applications for each family which best represents it in the current literature. However, to assure the operation of these applications, new requirements, far away from traditional services in fixed networks, appear in vehicular communications. To meet these demands, a number of communication technologies at level one/two are currently available. After explaining these, some of the most networking solutions at level three are described. Next, by means of an analysis which links applications, requirements and technologies, we give a vision of how each high level demand can be fulfilled by means of each technology. Since there is no any specific technology which can satisfy all the requirements, our opinion is that future vehicles will combine some of them in order to enable the deployment of different applications inside the vehicle.

Having the requirements and technologies in mind, vehicular network solutions can be designed by means of new communication architectures or integrating current ones. In this frame, our contribution should help in the design of both vehicular applications/services and technologies for tests scenarios, evaluations or commercial developments.

## REFERENCES

3G/UMTS evolution: towards a new generation of broadband mobile services [Manuel de logiciel]. (2006, December).

Adrisano, O., Verdone, R., & M., N. (2000, September). Intelligent transportation systems: The role of third-generation mobile radio networks. IEEECommunications Magazine, 38(9), 144-151.

Alexiou, A., Bouras, C., & Igglesis, V. (2004, November/ December). Performance evaluation of tcp over umts
transport channels. In International symposium on communicationsinterworking. Ottawa, Canada.

Alshaer, H., & Horlait, E. (2005). An optimized adaptive broadcast scheme for inter-vehicle communication. In IEEE vehicular technology conference. Stockholm, Sweden.

Ammoun, S., Nashashibi, F., & Laurgeau, C. (2006, September).Real-time crash avoidance system on crossroads based on 802.11 devices and GPS receivers. In IEEE intelligent transportation systems conference. Toronto, Canada.

Baldessari, R., Festag, A., & Lenardi, M. (2007, July). C2c-c consortium requirements for nemo route optimization
[Manuel de logiciel]. (IETF, draft-baldessari-c2ccc-nemoreq-01)

Baldessari, R., Festag, A., Matos, A., Santos, J., & Aguiar, R.(2006). Flexible connectivity management in vehicular communicationnetworks. In Proc. of the WIT 2004

Baldessari, R., Festag, A., Zhang,W., & Le, L. (2008). A manetcentric solution for the application of nemo in vanet using geographic routing. In Proc. of th weedev. Austria.

Bernardos, C. J., Soto, I., Calder´on, M., Boavida, F., & Azcorra, A. (2007). Varon: Vehicular ad hoc route optimisation for nemo. Comput. Commun., 30(8), 1765–1784.

Biswas, S., Tatchikou, R., & Dion, F. (2006, January). Vehicleto-vehicle wireless communication protocols for enhancinghighway traffic safety. IEEE Communications Magazine,44(1), 74-82.

Boukerche, A., Oliveira, H., Nakamura, E., Jang, K., & Loureiro, A. (2008, July). Vehicular ad hoc networks: A new challenge for localization-based systems. Computer Communications, Elsevier, 31(12), 2838-2849.

BREITENBERGER, S., GRBER, B., NEUHERZ, M., & KATES, R. (2004, July). Traffic information potential and
necessary penetration rates. Traffic engineering & control,45(11), 396–401.

Car-to-car communication consortium: http://www.car-tocar. org. (s. d.).

Chakeres, I., & Belding-Royer, M. (2004). AODV routing protocol implementation design. In Proceedings of the internationalworkshop on wireless ad hoc networking (WWAN).Tokyo, Japan.

Chatzigiannakis, I., Nikoletseas, E., & Spirakis, P. (2001). An efficient communication strategy for ad-hoc mobile networks. In 15th international conference on distributed computing
(disc). London, UK.

Chen, L., Steenstra, J., & K-S taylor. (2008, January).Geolocation-based addressing method for ipv6 addresses
(Patent). Qualcomm Incorporated.

Clausen, T., Jacquet, P., Laouiti, A., Muhlethaler, P., Qayyum, A., Viennot, L. (2001). Optimized link state routing protocol. In Proceedings of IEEE international multitopic conference INMIC. Pakistan.

Clifford Neuman, B. (1994). Scale in distributed systems. InReadings in distributed computing systems (pp. 463–489).IEEE Computer Society Press.

Devarapalli, V.,Wakikawa, R., Petrescu, A. and Thubert, P. (2005, January). Network mobility (NEMO) basic support protocol. (IETF RFC3963)

Duros, E., Dabbous, W., Izumiyama, H., Fujii, N., & Zhang, Y.(2001, March). A link-layer tunneling mechanism for unidirectional links. (IETF RFC3077)

Eddy, W., Ivancic, W., & Davis, T. (2007, December). Nemo route optimization requirements for operational use in aeronauticsand space exploration mobile networks. (IETF, draft-ietf-mext-aero-reqs-00)

ElBatt, T., Goel, S., Holland, G., Krishnan, H., & Parikh, J. (2006, September). Cooperative collision warning using dedicated short range wireless communications. In International conference on mobile computing and networking, international workshop on vehicular ad hoc networks. Los Angeles, USA.

Ernst, T., Montavont, N., Wakikawa, R., Ng, C., & Kuladinithi,K. (2007, July). Motivations and scenarios for using multiple interfaces and global addresses [Manuel de logiciel]. (IETF, draft-ietf-monami6-multihoming-motivation-scenario-02)

The eSafety initiative. (s. d.). http://www.esafetysupport.org/.

European telecommunications standards institute. (s. d.). Disponible sur http://www.etsi.org

Geonet project: http://www.geonet-project.eu. (s. d.).

Guo, M., M-H., A., & Zegura, E.-W. (2005). V3: A vehicle-tovehicle live video streaming architecture. In Percom (p. 171-180).

Han, M., Moon, S., Lee, Y., Jang, K., & Lee, D. (2008, April). Evaluation of MoIP quality overWiBro. In Passive and active measurement conference. Cleveland, USA.

Hattori, G., Ono, C., Nishiyama, S., & Horiuchi, H. (2004, January). Implementation and evaluation of message delegation middleware for ITS application. In International symposium on applications and the internet workshops. Tokyo, Japan.

Hoh, B., Gruteser, M., Xiong, H., & Alrabady, A. (2006, October-December). Enhancing security and privacy in
traffic-monitoring systems. IEEEPervasive Computing, 5(4), 38-46.

The Intelligent Transportation System. (s. d.). http://www.its.dot.gov/its overview.htm.

J. Lorchat and K. Uehara. (2006, July). Optimized Inter-Vehicle Communications Using NEMO and MANET (Invited Paper). (The Second International Workshop on Vehicle-to-Vehicle Communications 2006 (V2VCOM 2006))

Jiang, M., Li, J., & Tay, Y. (2001). Cluster based routing protocol (CBRP) (Rapport technique). IETF. (Internet draft) Johnson, D., Perkins, C., & Arkko, J. (2004, June). Mobility support in ipv6 [Manuel de logiciel]. (IETF RFC 3775)

Khaled, Y., Ducourthial, B., & Shawky, M. (2005-Spring). IEEE 802.11 performances for inter-vehicle communication networks. In Proc. of th 61st IEEE semianual vehicular technology conference VTC. Stockholm, Sweden.

Khaled, Y., Ducourthial, B., & Shawky, M. (july, 2007). A usage oriented taxonomy of routing protocols in vanet. In Proceedings of 1st ubiroads workshop with ieee giis. Marrakech, Morroco.

Kiess, W., Rybicki, J., & Mauve, M. (2007, February/March). On the nature of inter-vehicle communications. In Workshop on mobile ad-hoc networks. Bern, Switzerland.

King, T., Füßler, H., Transier, M., & Effelsberg, W. (2006, 03). Dead-Reckoning for Position-Based Forwarding on Highways. In Proc. of the 3rd international workshop on intelligent transportation (WIT 2006) (p. 199-204). Hamburg, Germany.

Kosch, T. (2002). Technical concept and prerequisites of carto-car communication (Rapport technique). BMW Group Research and Technology.

Landman, J., & Kritzinger, P. (2005). Delay analysis of downlink IP traffic on umts mobile networks. Perform. Eval., 62(1-4), 68–82.

Larsson, C., Eriksson, M., Mitsuya, K., Tasaka, K., & Kuntz, R.(2008, July). Flow distribution rule language for multi-access nodes [Manuel de logiciel]. (IETF, draft-larsson-mext-flowdistribution-rules-00)

Li, Q., & Rus, D. (2000). Sending messages to mobile users in disconnected ad-hoc wireless networks. In 6th annual international conference on mobile computing and networking (MOBICOM).

Maihöfer, C. (2nd quarter 2004). A survey of geocast routing protocols. IEEE Communications Surveys and Tutorials, 6.

Masini, B., Fontana, C., & Verdone, R. (2004, October). Provision of an emergency warning service through gprs: Performance evaluation. In IEEE intelligent transportation systems conference. Washington, USA.

Mauve, M., Widmer, J., & Hartenstein, H. (2001, November/December). A survey on position-based routing in mobile ad hoc networks. IEEE Network Magazine.

Meier, R., Hughes, B., Cunningham, R., & Cahill, V. (2005). Towards real-time middleware for applications of vehicular ad hoc networks. In Ifip wg 6.1 international conference, distributed applications and interoperable systems. Oslo, Norway.

Naumov, V., Baumann, R., & Gross, T. (2006, May). An evaluation of inter-vehicle ad hoc networks based on realistic vehicular traces. In ACM international symposium on mobile ad hoc networking and computing. Florence, Italia.

Ng, C., Ernst, T., Paik, E., & Bagnulo, M. (2007, October). Analysis of multihoming in network mobility support. (IETF, RFC4980).

Ng, C., Thubert, P., Watari, M., & Zhao, F. (2007, July). Network mobility route optimization problem statement (IETF RFC4888).

Nolte, T., Hansson, H., & Lo Bello, L. (2005, September). Automotive communications - past, current and future. In IEEE internationalconference on emerging technologies and factory automation. Catania, Italy.

Qureshi, A., Carlisle, J., & Guttag, J. (2006, October). Tavarua: Video streaming with wwan striping. In Acm multimedia 2006. Santa Barbara, CA.

Santa, J., & Gomez-Skarmeta, A. (2008, July). Architecture andevaluation of a unified V2V and V2I communication system based on cellular networks. Elsevier Computer Communications, 31(12), 2850-2861.

Sawant, H., Tan, J., Yang, Q., & Wang, Q. (2004, October). Using bluetooth and sensor networks for intelligent transportation systems. In Ieee international conference on intelligent transportation systems. Washington DC, USA.

Soliman, H., Montavont, N., Fikouras, N., & Kuladinithi, K. (2007, November). Flow bindings in mobile ipv6 and nemo basic support [Manuel de logiciel]. (IETF, draft-solimanmonami6-flow-binding-05)

Sugiura, A., & Dermawan, C. (2005, September). In traffic jam IVC-RVC system for ITS using bluetooth. IEEE Transactions on Intelligent Transportation Systems, 6(3), 302- 313.

Tsukada, M., Mehani, O., & Ernst, T. (2008, March). Simultaneous usage of NEMO and MANET for vehicular
communication. In WEEDEV 2008: 1st workshop on experimental evaluation and deployment experiences on vehicular networks in conjonction with TRIDENTCOM 2008, innsbruck, austria, march 18, 2008. Disponible sur
http://hal.inria.fr/inria-00265652/

Ueki, J., Tasaka, S., Hatta, Y., & Okada, H. (2005). Vehicular collisionavoidance support system (vcass) by inter-vehicle communications for advanced its. IEICE Transactions on Fundamentals of Electronics, Communications and Computer Sciences, E88-A(7), 1816–1823.

Vare, J., Syrjarinne, J., & Virtanen, K.-S. (2004). Geographical positioning extension for IPv6. In Proc. of the icn. Guadeloupe.

Wakikawa, R., Clausen, T., McCarthy, B., & Petrescu, A. (2007, July). Manemo topology and addressing architecture. (IETF, draft-wakikawa-manemoarch-00)

Wakikawa, R., Ernst, T., Nagami, K., & Devarapalli, V. (2008, January). Multiple care-of addresses registration. (IETF, draft-ietf-monami6-multiplecoa-05).

Wakikawa, R., Okada, K., Koodli, R., Nilsson, A., & Murai, J. (2005, September). Design of Vehicle Network: Mobile Gatewayfor MANET and NEMO Converged Communication. (The Second ACM International Workshop on Vehicular Ad Hoc Networks (VANET 2005))

Wakikawa, R., Thubert, P., Boot, T., Bound, J., & McCarthy,B. (2007, July). Problem statement and requirements for manemo [Manuel de logiciel]. (IETF, draft-mccarthymanemo-configuration problems-01)

Wewetzer, C., Caliskan, M., Meier, K., & Luebke, A. (2007, June). Experimental evaluation of umts and wireless lan for inter-vehicle communication. In International conference its telecommunications. Sophia Antipolis, France.

Yousefi, S., Bastani, S., & Fathy, M. (2007, February). On the performance of safety message dissemination in vehicular ad hoc networks. In European conference on universal multiservice networks. Toulouse, France.

Zhao, J.,&Cao, G. (2006). VADD: Vehicle-assisted data delivery in vehicular ad hoc networks. In 25th conference on computer communications *(INFOCOM)*. Barcelona, Spain.